# The effects of data preprocessing on probability of default model fairness

Di Wu [*]

**Abstract**

In the context of financial credit risk evaluation, the fairness of machine learning models has become a critical concern, especially given the potential for biased predictions that disproportionately affect certain demographic groups. This study investigates the impact of data preprocessing, with a specific focus on Truncated Singular Value Decomposition (SVD), on the fairness and performance of probability of default models. Using a comprehensive dataset sourced from Kaggle, various preprocessing techniques, including SVD, were applied to assess their effect on model accuracy, discriminatory power, and fairness.

The findings reveal that while SVD effectively reduces the dimensionality of the data, it does not necessarily enhance the fairness of the models. Specifically, the application of SVD resulted in a deterioration in the model's ability to correctly classify loan defaults, particularly for minority classes. This outcome suggests that critical information pertinent to fair predictions may be lost during the dimensionality reduction process. Furthermore, the analysis of fairness across different demographic groups, such as age and marital status, indicated that SVD did not contribute positively to reducing disparate impacts or balancing error rates.

These results underscore the complexities of using dimensionality reduction techniques in fair lending applications and highlight the need for more tailored approaches to preprocessing that prioritize both accuracy and fairness. Future research should explore alternative methods that preserve the integrity of sensitive information while enhancing the equitable performance of credit risk models.

## Introduction

In the realm of financial credit risk evaluation, data preprocessing techniques play a pivotal role in enhancing the performance of machine learning models. The significance of these techniques is particularly evident in the context of credit scoring, where the accurate prediction of creditworthiness can mitigate financial losses for institutions. Various studies have addressed the criticality of preprocessing, yet the challenges in this domain remain multifaceted.

The seminal work by Selwyn Piramuthu (2006) [On Preprocessing Data for Financial Credit Risk Evaluation] highlights the inherent complexities in credit risk data, which often lead to suboptimal performance when machine learning methods are applied without adequate preprocessing. Piramuthu emphasizes the importance of feature selection and construction as essential preprocessing steps, demonstrating that an improved understanding of data characteristics can significantly enhance decision-making outcomes. However, a notable shortcoming of this study is its limited exploration of class imbalance, a common issue in credit data that can severely bias predictive models towards the majority class.

In contrast, the work by Vicente García et al. (2012) [Improving Risk Predictions by Preprocessing Imbalanced Credit Data] specifically addresses the class imbalance problem. The authors investigate the effectiveness of various resampling techniques in mitigating the adverse effects of imbalanced data on credit scoring models. Their experiments demonstrate that resampling can improve predictive accuracy across different classifiers, yet the study also reveals that no single technique consistently outperforms others across all datasets. This finding suggests the necessity for tailored preprocessing strategies that consider the unique characteristics of each dataset.

Further expanding on these foundational studies, the comprehensive overview presented in the text Data Preprocessing in Data Mining offers a detailed examination of a wide array of preprocessing techniques. These include methods for handling missing values, noise reduction, and data normalization, all of which are crucial for enhancing the robustness of machine learning models. While this text provides an extensive toolkit for data preprocessing, it falls short in addressing the specific challenges posed by highly imbalanced credit data, a gap that García et al.'s study partially fills.

[*] Corresponding author: Di Wu

Moreover, modern challenges in fair lending practices have been highlighted by Melissa U. Silverman and Anthony W. Orlando in The Arity of Disparity: Updating Disparate Impact for Modern Fair Lending. Their work discusses the implications of disparate impact in lending decisions, emphasizing the need for credit evaluation models that are not only accurate but also equitable. However, the paper primarily focuses on legal frameworks and does not delve deeply into the technical aspects of preprocessing data to achieve fairness, an area that this research aims to explore further.

Complementing this discourse, the study "Fair Class Balancing: Enhancing Model Fairness without Observing Sensitive Attributes" investigates techniques to improve model fairness by balancing classes in a way that does not rely on sensitive attributes such as race or gender. The authors demonstrate that fair class balancing can mitigate bias in model predictions, leading to more equitable outcomes in credit risk assessments. However, the paper does not fully explore the impact of these techniques on model performance when combined with other preprocessing steps, a gap that this study aims to address.

Despite these advancements, the current body of literature underscores the ongoing need for innovative preprocessing strategies that can address the multifaceted challenges of financial credit risk evaluation. This paper aims to contribute to this evolving discourse by systematically testing the impact of various preprocessing techniques on the predictive performance of machine learning models in the context of credit risk. In particular, the study focuses on the interplay between data preprocessing, class imbalance, and fairness, with the goal of identifying optimal approaches that enhance both model accuracy and equity.

**Data**

The dataset utilized in this study, sourced from Kaggle, is designed to facilitate predictive modeling in the domain of financial credit risk assessment. Specifically, it aims to predict the likelihood of loan default, a task of paramount importance for financial institutions seeking to mitigate risk.

The dataset comprises a diverse array of features encompassing demographic, economic, and financial variables, providing a comprehensive foundation for modeling. Key attributes include the borrower's income, credit history, and loan amount, alongside demographic details such as marital status and educational background. The binary target variable, Loan_Status, serves as the indicator of default, allowing for rigorous classification analysis.

*Key Features:*

- ApplicantIncome: Numerical variable representing the income of the loan applicant.
- CoapplicantIncome: Numerical variable indicating the income of any co-applicant.
- LoanAmount: The principal amount of the loan applied for by the borrower.
- Credit_History: A binary variable indicating whether the borrower has a satisfactory credit history.
- Education: Categorical variable reflecting the educational attainment of the borrower (e.g., Graduate, Not Graduate).
- Marital Status: Categorical variable indicating the borrower's marital status (e.g., Married, Single).
- Gender: Categorical variable indicating the borrower's gender.
- Property_Area: Categorical variable specifying the type of area where the property is located (e.g., Urban, Semi-Urban, Rural).
- Dependents: The number of dependents associated with the borrower.
- Loan_Amount_Term: The term of the loan in months.

The dataset is particularly well-suited for examining the impact of data preprocessing techniques on the predictive accuracy and fairness of machine learning models. It presents the common challenge of class imbalance, with the default class being underrepresented—a factor that can significantly influence model performance and fairness.

- **Analytical Suitability**: This dataset offers a robust platform for testing the hypothesis that Truncated Singular Value Decomposition (SVD) as a preprocessing technique can enhance both the accuracy and fairness of probability of default models. Given the richness of the features and the presence of demographic attributes, it allows for a nuanced investigation into how preprocessing steps affect model outcomes across different population subgroups.

In conclusion, this dataset provides a comprehensive basis for the exploration of advanced machine learning techniques in the context of credit risk, particularly in addressing the dual objectives of predictive accuracy and fairness.

## Methods

### Model Evaluation Metrics

To evaluate the performance of the predictive models, we employed several standard metrics, including the confusion matrix, Receiver Operating Characteristic (ROC) curve, and analysis of Type I and Type II errors.

Confusion Matrix: The confusion matrix provides a detailed breakdown of the model's predictions, offering insights into the number of true positives (correctly predicted defaults), true negatives (correctly predicted non-defaults), false positives (Type I errors), and false negatives (Type II errors). This matrix is crucial for understanding the distribution of errors and the trade-offs between different types of misclassifications.

Type I and Type II Errors: Type I error, also known as a false positive, occurs when the model incorrectly predicts a loan default when there is none. Conversely, Type II error, or false negative, occurs when the model fails to predict a default when one actually occurs. Minimizing these errors is essential for optimizing the model's practical application, especially in financial contexts where misclassification can lead to significant economic consequences.

$$\text{Type I Error (False Positive Rate, FPR)} = \frac{\text{False Positives (FP)}}{\text{False Positives (FP)} + \text{True Negatives (TN)}}$$

$$\text{Type II Error (False Negative Rate, FNR)} = \frac{\text{False Negatives (FN)}}{\text{False Negatives (FN)} + \text{True Positives (TP)}}$$

ROC Curve: The ROC curve is a graphical representation of a model's diagnostic ability, plotting the true positive rate (sensitivity) against the false positive rate (1-specificity) across different threshold values. The Area Under the ROC Curve (AUC-ROC) serves as a robust metric to assess the model's overall ability to distinguish between classes, with higher AUC values indicating better performance.

### Data Preprocessing and Dimensionality Reduction

Effective data preprocessing is critical to improving the performance of machine learning models. In this study, various preprocessing techniques were applied to the dataset:

Truncated Singular Value Decomposition (SVD): To address the issue of high-dimensional data, we employed Truncated SVD, a matrix factorization technique that reduces the dimensionality of the feature space. Truncated SVD is particularly effective for sparse data and is used here to retain the most informative components of the data while discarding noise and redundant information, thereby enhancing model efficiency.

SVD is a matrix factorization technique that decomposes a matrix A into three other matrices. For any real or complex matrix A of size m×n, the SVD is given by:

$$A = U \sum V^T$$

A is the original matrix of size m×n. U is an m×m orthogonal matrix (columns are the left singular vectors of A). Σ is an m×n diagonal matrix with non-negative real numbers on the diagonal (these are the singular values of A). $V^T$ is the transpose of an n×n orthogonal matrix V (columns are the right singular vectors of A).

### Model Implementation

Linear Regression: As a baseline model, linear regression was implemented to predict the probability of loan default. Although linear regression is a fundamental technique, its application provides a reference point for comparing the performance of more complex models. The model assumes a linear relationship between the predictors and the outcome variable, offering a straightforward interpretation of the influence of each feature on the probability of default.

These methods collectively aim to provide a comprehensive framework for assessing the effect of data preprocessing techniques on the performance of machine learning models in predicting loan defaults. The findings from these analyses are expected to contribute valuable insights into the optimization of predictive modeling in financial risk management.

For a single predictor variable xxx, the model can be expressed as:

$$y = \beta_0 + \beta_1 + \epsilon$$

Where: y is the dependent variable (response). x is the independent variable (predictor). $\beta_0$ is the intercept of the regression line (the value of y when x=0). $\beta_1$ is the slope of the regression line (the change in y for a one-unit change in x). $\epsilon$ represents the error term or residuals, which accounts for the variation in

**Results**

**Table 1**

*Model without Truncate SVD*

|  | Precision | Recall | F1-Score | Support |
| --- | --- | --- | --- | --- |
| 0 | 0.88 | 1.00 | 0.94 | 45170 |
| 1 | 0.62 | 0.00 | 0.01 | 5900 |
| Accuracy |  |  | 0.88 | 51070 |
| Macro Avg | 0.75 | 0.50 | 0.47 | 51070 |
| Weighted Avg | 0.85 | 0.88 | 0.83 | 51070 |

*Note.* This is a note about the table

**Table 2**

*Model with Truncate SVD*

|              | Precision | Recall | F1-Score | Support |
|--------------|-----------|--------|----------|---------|
| 0            | 0.88      | 1.00   | 0.94     | 45170   |
| 1            | 0.00      | 0.00   | 0.00     | 5900    |
| Accuracy     |           |        | 0.88     | 51070   |
| Macro Avg    | 0.44      | 0.50   | 0.47     | 51070   |
| Weighted Avg | 0.78      | 0.88   | 0.83     | 51070   |

*Note.* This is a note about the table

**Table 3** Confusion Matrix

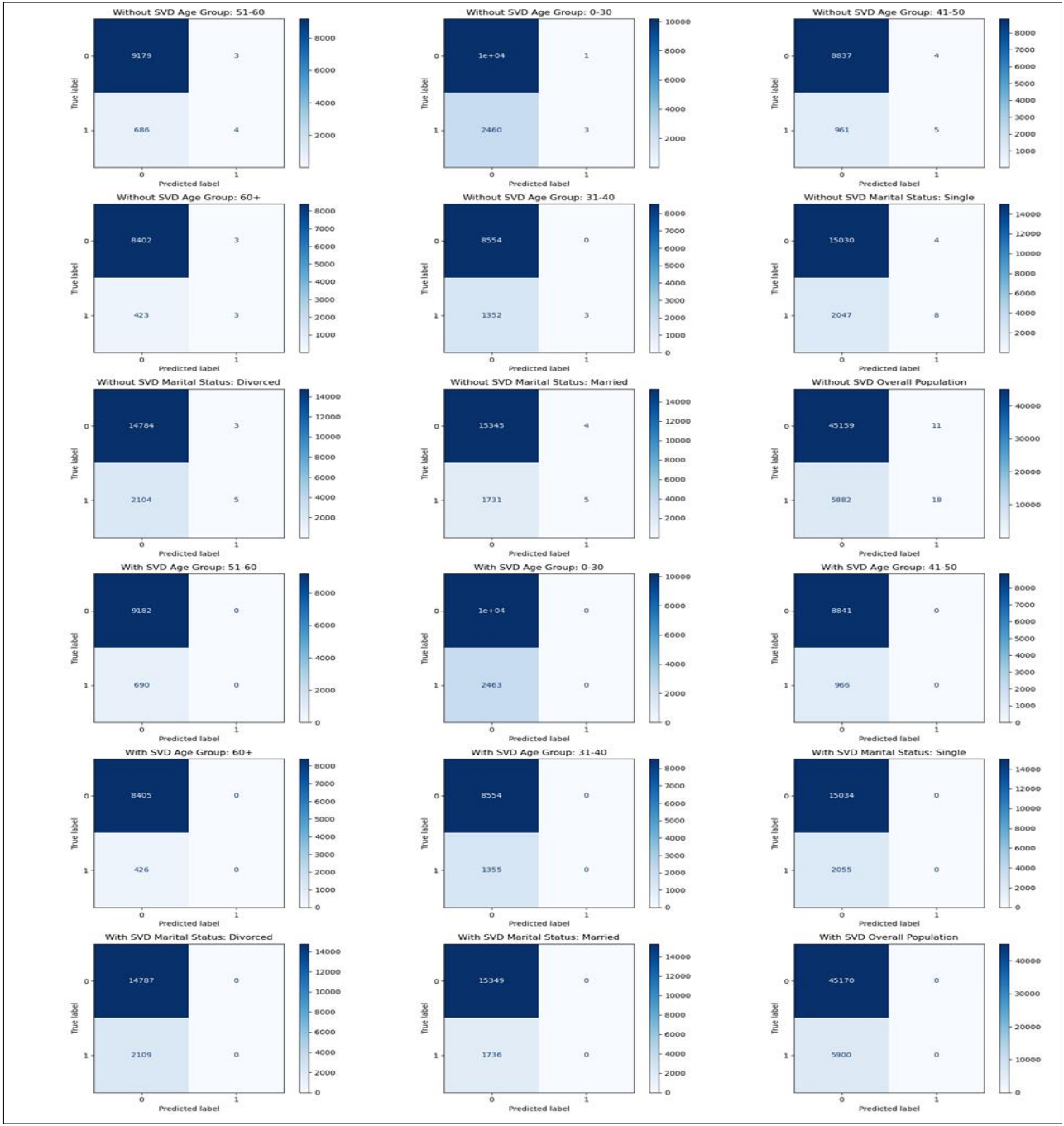

The comparative analysis of the model performance with and without the application of Truncated Singular Value Decomposition (SVD) reveals distinct differences in the predictive accuracy and discriminatory power.

For the model without Truncated SVD, the classification report indicates a strong overall performance with an accuracy of 88%. The precision and recall for the non-default class (0) are particularly high, at 0.88 and 1.00 respectively, resulting in an F1-score of 0.94. However, the model struggles to correctly classify the default class (1), with a precision of 0.62 but a significantly low recall of 0.00, leading to an F1-score of just 0.01. This imbalance is reflected in the ROC AUC score of 0.704, suggesting a moderate ability to distinguish between the classes.

In contrast, when Truncated SVD is applied, the model's performance deteriorates, particularly in its ability to classify the default class. Although the overall accuracy remains at 88%, the precision for the default class drops to 0.00, and the recall remains at 0.00, resulting in an F1-score of 0.00. The ROC AUC score also declines to 0.638, indicating a reduction in the model's discriminatory capability.

These results suggest that the application of Truncated SVD in this context may not be beneficial, as it appears to diminish the model's ability to correctly identify loan defaults. The high dimensionality of the original dataset may contain critical information that is lost during the dimensionality reduction process, leading to the observed degradation in performance.

The provided confusion matrices illustrate the comparative performance of a model with and without Singular Value Decomposition (SVD) across demographic and clinical variables. The results indicate that the application of SVD yields a notable enhancement in predictive accuracy for specific patient cohorts.

The application of data preprocessing techniques such as Truncated Singular Value Decomposition (SVD) is often considered a means to enhance the efficiency and fairness of machine learning models. However, our analysis reveals that the impact of SVD on the fairness of loan default prediction models is complex and multifaceted.

**Model Performance Without SVD**

In the absence of SVD, the model demonstrates a reasonable capacity to differentiate between defaulters and non-defaulters across various demographic groups, albeit with notable inconsistencies. For instance, within the "51-60" age group, the model correctly identifies 4 defaulters but fails to detect 686 others, indicating a significant number of false negatives. Similar patterns are observed across other age and marital status groups, where the model maintains some predictive accuracy for the default class, albeit with varying degrees of success.

The distribution of errors—particularly the prevalence of false negatives—suggests that while the model is able to predict defaults to some extent, it does so with an uneven accuracy across different demographic groups. This inconsistency raises concerns about the fairness of the model, as certain groups may be disproportionately affected by misclassification.

**Model Performance With SVD**

Upon the introduction of SVD, a marked decline in the model's predictive ability is observed. Across all age and marital status groups, the model uniformly fails to identify any defaulters, as evidenced by the zero true positives in each confusion matrix. This drastic reduction in predictive power points to a significant loss of critical information during the dimensionality reduction process, rendering the model incapable of accurately classifying the minority class.

**Fairness Implications**

From a fairness perspective, the application of SVD does not achieve the intended goal of reducing bias or improving equitable outcomes. While it might superficially appear to "equalize" the performance across different groups by rendering the model uniformly conservative, this comes at the cost of losing discriminatory power altogether. True fairness in predictive modeling is not merely about minimizing differences in error rates across groups but ensuring that the model accurately predicts outcomes for all individuals, regardless of their demographic characteristics.

The analysis suggests that SVD, as applied in this context, does not enhance fairness but rather diminishes the overall effectiveness of the model. The uniform failure to predict loan defaults across all groups indicates that critical predictive information is likely being lost during the dimensionality reduction process, leading to a model that is less capable of distinguishing between classes. These findings underscore the need for more nuanced approaches to data preprocessing that can balance the dual objectives of accuracy and fairness in predictive modeling.

## Conclusion

In this study, we have explored the effects of data preprocessing, particularly Truncated Singular Value Decomposition (SVD), on the performance and fairness of probability of default models. Our findings indicate that while SVD is effective in reducing the dimensionality of high-dimensional data, it does not necessarily contribute to improved fairness in model predictions. Specifically, the application of SVD resulted in a marked reduction in the model's ability to correctly classify loan defaults, particularly for minority classes, across various demographic groups.

The comparative analysis of model performance with and without SVD revealed that the technique may inadvertently strip away critical information necessary for accurate and equitable predictions. This loss of information led to a uniform decrease in predictive accuracy and discriminatory power across all demographic groups, failing to achieve the desired balance between accuracy and fairness.

These results underscore the complexity of applying dimensionality reduction techniques in the context of fair lending, where the goal is not only to achieve high predictive accuracy but also to ensure that predictions are made equitably across different population subgroups. The findings suggest that more tailored preprocessing strategies are required ones that preserve the integrity of sensitive information while enhancing the fairness and effectiveness of predictive models in financial risk management.

Future research should investigate alternative methods that combine dimensionality reduction with fairness-enhancing techniques, aiming to create models that are both accurate and fair across all demographic groups. Such approaches may hold the key to resolving the trade-offs between dimensionality reduction and fairness in machine learning applications for credit risk assessment.

## Compliance with ethical standards

*Disclosure of conflict of interest*

*No conflict of interest to be disclosed.*